\documentclass[runningheads]{llncs}
\usepackage{graphicx}
\usepackage[dvipsnames]{xcolor}
\usepackage{amsmath,amssymb,booktabs,microtype}

\usepackage[frozencache=true,cachedir=_minted-output]{minted}

\setminted{fontsize=\footnotesize}

\usepackage{hyperref}

\renewcommand{\orcidID}[1]{\texorpdfstring{~\href{https://orcid.org/#1}{\raisebox{-0.3ex}{\protect\includegraphics[height=1em]{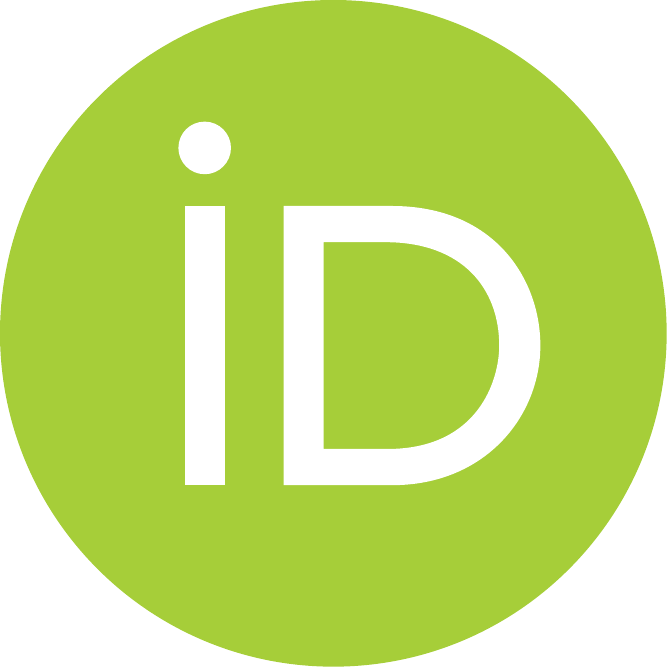}}}}{}}

\begin{document}
\emergencystretch=2dd
\hyphenation{web-mias mias}
\setlength{\abovecaptionskip}{.3\abovecaptionskip}
\addtolength{\textheight}{\baselineskip}
\renewcommand{\baselinestretch}{0.98}
\title{WebMIaS on Docker\thanks{The second author's work was graciously funded by the South Moravian Centre for International Mobility as a part of the Brno Ph.D.\ Talent project.}}
\subtitle{Deploying Math-Aware Search in a Single Line of Code}
\titlerunning{WebMIaS on Docker: Math-Aware Search in a Single Line of Code}
\author{Dávid~Lupták\orcidID{0000-0001-9600-7597}
\and Vít Novotný\orcidID{0000-0002-3303-4130}
\and Michal~Štefánik\orcidID{0000-0003-1766-5538}
\and Petr Sojka\orcidID{0000-0002-5768-4007}}
\authorrunning{D. Lupták, V. Novotný, M. Štefánik, and P. Sojka}
\institute{Faculty of Informatics, Masaryk University,
Brno, Czech Republic\\
\email{\{dluptak,witiko,stefanik.m\}@mail.muni.cz},
\email{sojka@fi.muni.cz}\\
\url{https://mir.fi.muni.cz/}
}

\maketitle              
\begin{abstract}
\vspace*{-1.5\baselineskip}

Math informational retrieval (MIR) search engines are absent in the wide-spread production use, even though documents in the STEM fields contain many mathematical formulae, which are sometimes more important than text for understanding.
We have developed and open-sourced the WebMIaS MIR search engine that has been successfully deployed in the European Digital Mathematics Library (EuDML).
However, its deployment is difficult to automate due to the complexity of this task. Moreover, the solutions developed so far to tackle this challenge are imperfect in terms of speed, maintenance, and robustness.
In this paper, we will describe the virtualization of WebMIaS using Docker that solves all three problems and allows anyone to deploy containerized WebMIaS in a single line of code.
The publicly available Docker image will also help the community push the development of math-aware search engines in the ARQMath workshop series.

\end{abstract}

\section{Introduction}
\label{sec:1:intro}
\vspace*{-.5\baselineskip}

Searching for math formulae does not appear as a task for search engines at first glance.
Text retrieval is dominant among search engines, while math-awareness is a specialized area in the field of information retrieval:
\href{https://link.springer.com/}{Springer's} \LaTeX\ Search,
the MathWebSearch of \href{https://zbmath.org/formulae/}{zbMATH Open} (formerly known as Zentralblatt MATH), and the
Math Indexer and Searcher (MIaS) of the \href{https://eudml.org/search}{European Digital Mathematics Library (EuDML)}
are all examples of systems with math-aware search deployed in production.
Our MIaS search engine~\cite{mir:MIaSNTCIR-11short}
runs on the industry-grade, robust, and highly-scalable full-text search engine \href{https://lucene.apache.org/}{Apache Lucene} with our own preprocessing of mathematical formulae.
\begin{figure}[htb]
    \centering
    \includegraphics[width=.9\textwidth]{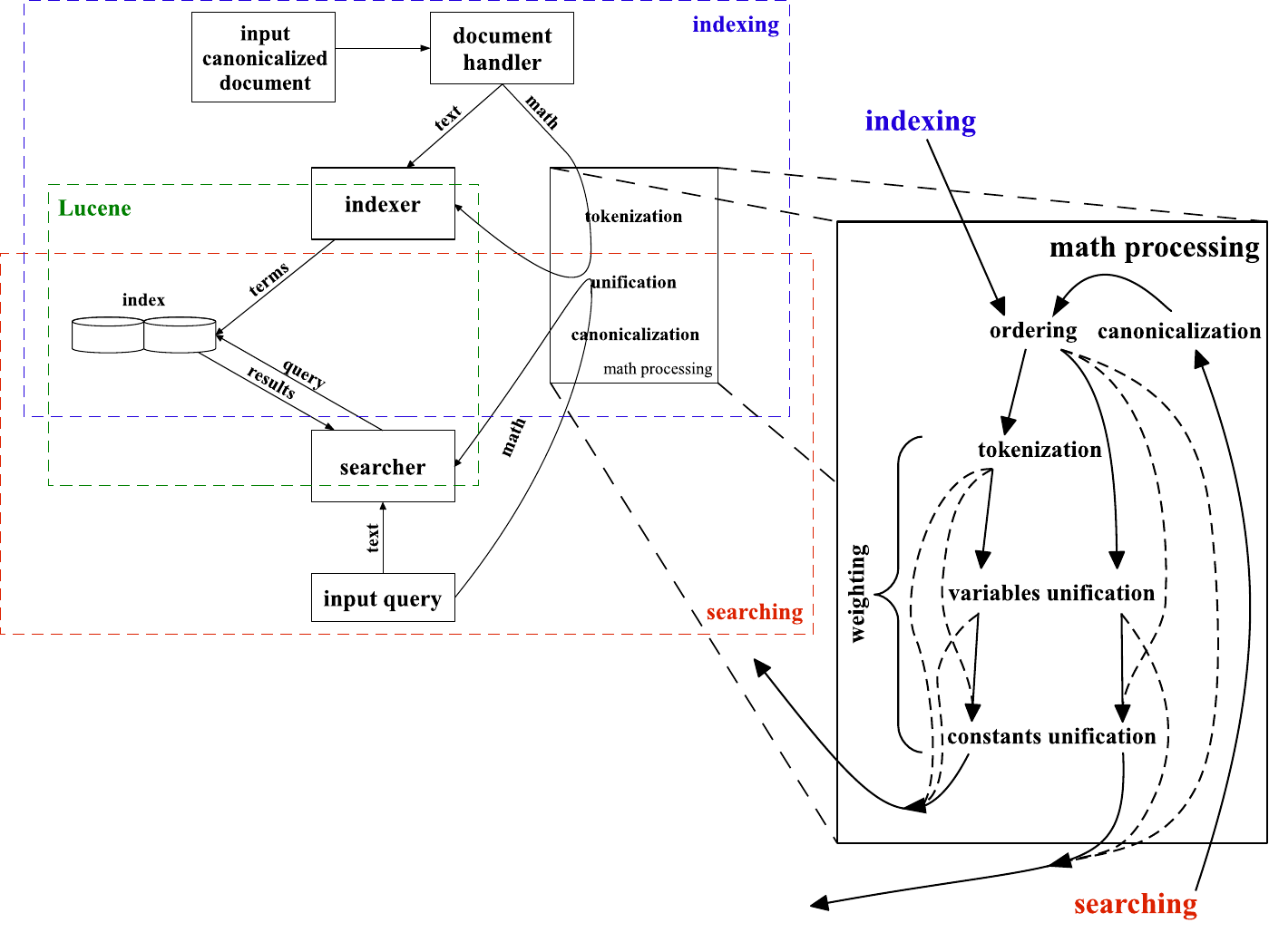}
    \vspace*{-.7\baselineskip}
    \caption{The architecture of MIaS with indexing and searching phases overlapping over \href{https://lucene.apache.org/}{Lucene} index. Besides standard text processing,
    the math input from indexing (a document) and searching (a query) stage is canonicalized, ordered, tokenized, and unified, afterward returned back to the indexer and searcher module, respectively.}
    \label{fig:system-architecture}
    \vspace*{-.5\baselineskip}
\end{figure}
The text is tokenized and stemmed to unify inflected word forms whereas math is expected to be in \href{https://www.w3.org/TR/MathML3/}{the MathML format}%
, which is
then canonicalized, ordered, tokenized, and unified, see Figure~\ref{fig:system-architecture}.

\begin{figure}[htb]
    \centering
    \includegraphics[width=.9\textwidth]{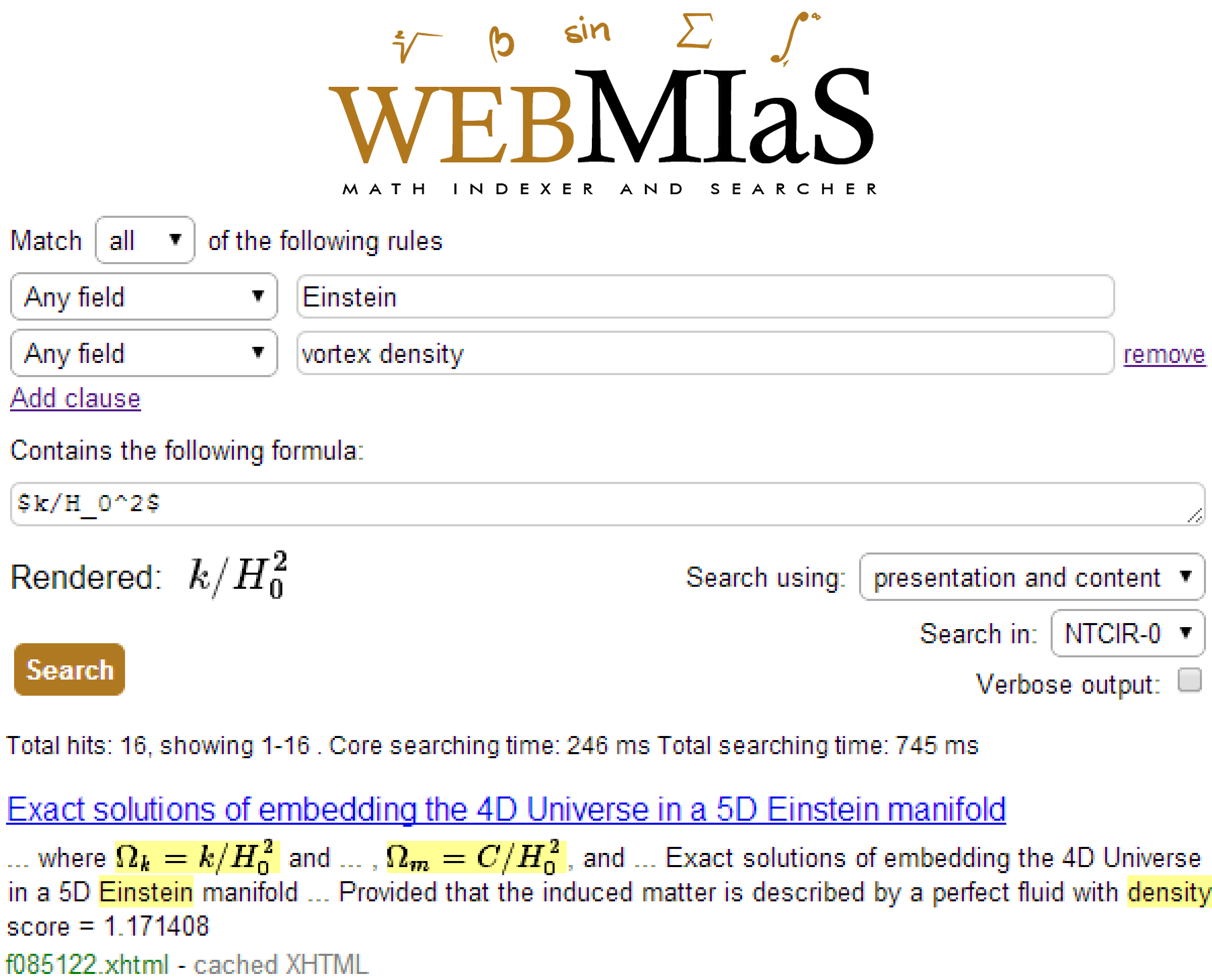}
    \caption{Searching text and formulae with a single mixed query in WebMIaS.}
    \label{fig:webmias-ui}
    \vspace*{-.5\baselineskip}
\end{figure}

To provide a web user interface for MIaS, we have developed and
open-sourced
the WebMIaS~\cite{mir:webmias2014short,mir:MIaSNTCIR-11short} search engine.
In WebMIaS, users can input their mixed queries in a combination of text and math with a native support for \LaTeX{} and MathML.
Matches are conveniently highlighted in the search results.
The user interface of WebMIaS is shown in Figure~\ref{fig:webmias-ui}.

Although the (Web)MIaS system has been deployed in the European Digital Mathematics Library (EuDML) already, the complicated deployment process might be an obstacle for a more wide-spread deployment to other digital mathematics libraries that avail of or can extend to the MathML markup.
To solve this problem, we will describe the virtualization of WebMIaS using Docker~\cite{boettiger2015introduction} that allows anyone to deploy WebMIaS in a single line of code.
Whether you have an open-access repository such as \href{https://duraspace.org/dspace/}{DSpace}, or just a number of mathematical documents, you can benefit from the math-aware search provided by WebMIaS.
For testing, we also provide the MREC dataset~\cite{dml:liska2011short}.

In the rest of our paper, we will describe our deployment process in Section~\ref{sec:2:deployment}, evaluate the speed and quality of WebMIaS in Section~\ref{sec:3:evaluation}, and conclude in Section~\ref{sec:4:conclusion}.

\section{Deployment process description}
\label{sec:2:deployment}
\vspace*{-.5\baselineskip}

All modules of the MIaS system are Java projects, so users first need to
1)~install the Java environment prerequisites and then 2)~build the respective system modules.
The next step in the process is to 3)~index a dataset of mathematical documents using the command-line interface of MIaS.
Finally, the users can 4)~run \href{https://tomcat.apache.org/}{Apache Tomcat} with the WebMIaS servlet as a user interface.

Over the years, we have attempted to automate the above steps into running a single \href{https://www.gnu.org/software/make/}{Makefile} or \href{https://jupyter.org/}{Jupyter Notebook}. However, these solutions were slow, fragile, and hard to maintain. We propose a better solution using lightweight virtualization via \href{https://www.docker.com/}{Docker} with instant deployment, a short but powerful Dockerfile configuration, and a complete workflow that automates all the steps of the deployment process.
Moreover, \href{https://github.com/features/actions}{GitHub Actions} provide continuous integration and automate the publishing of Docker images to \href{https://hub.docker.com/}{Docker Hub}.

\begin{figure}[tb]
    \leavevmode\kern-0.15\textwidth
    \includegraphics[width=1.3\textwidth]{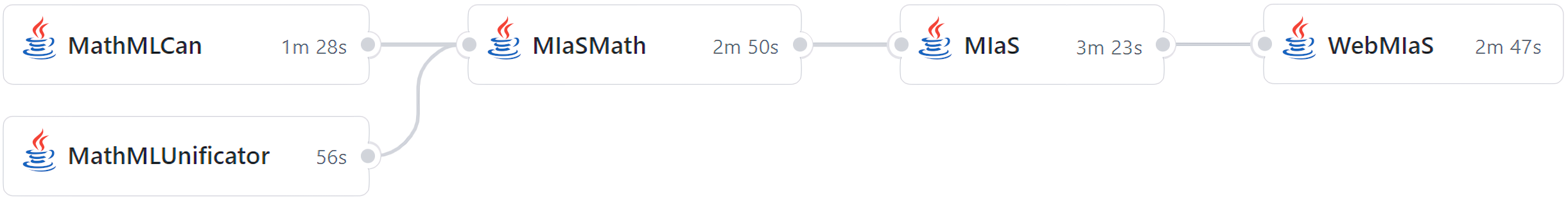}
    \caption{The continuous integration of \emph{WebMIaS} and the build times of the respective packages: \emph{MathMLCan} canonicalizes different \href{https://www.w3.org/TR/MathML3/}{MathML} encodings of equivalent formulae. \emph{MathMLUnificator} generalizes distinct mathematical formulae so that they can be structurally unified. \emph{MIaSMath} adds math processing capabilities to \href{https://lucene.apache.org/}{Lucene} or \href{https://solr.apache.org/}{Solr}. \emph{MIaS} indexes text with math in Lucene/Solr-based full-text search engines. Finally, \emph{WebMIaS} provides a web interface for \emph{MIaS}.}
    \label{fig:CI}
\end{figure}

Both MIaS and WebMIaS are containerized into separate Docker images named
\href{https://hub.docker.com/r/miratmu/mias/tags}{\texttt{miratmu/mias}} and 
\href{https://hub.docker.com/r/miratmu/webmias/tags}{\texttt{miratmu/webmias}}, respectively.
This allows users to run both the indexing and the retrieval without a specific configuration of the environment.
Resolving the dependencies and building all modules is up to the continuous integration workflow (see Figure~\ref{fig:CI}), and users receive Docker images with everything prebuilt.
After downloading a dataset to the working directory, users can index the \texttt{dataset} directory into the \texttt{index} directory using MIaS, see Listing~\ref{listing:1}.
\begin{listing}[t]
\begin{minted}[
  linenos,
  breaklines,
%  frame=lines,
  framesep=.02mm,
]{bash}
$ wget https://mir.fi.muni.cz/MREC/MREC2011.4.439.tar.bz2
$ mkdir dataset ; tar xj -f MREC2011.4.439.tar.bz2 -C dataset
$ docker run -v "$PWD"/dataset:/dataset:ro -v "$PWD"/index:/index:rw --rm miratmu/mias
$ docker run -v "$PWD"/dataset:/dataset:ro -v "$PWD"/index:/index:ro --rm --name webmias -d -p 127.0.0.1:8888:8080 miratmu/webmias
\end{minted}
\caption{\small Downloading and indexing the MREC2011.4 dataset for WebMIaS (lines 1--3), and deploying WebMIaS in a single line (n.\,4) of code.}
\label{listing:1}
\end{listing}

\begin{table}[b!]
\caption{The linear indexing speed on the MREC dataset using 448G of RAM, and eight Intel Xeon\texttrademark{} X7560 2.26\,GHz CPUs.}
\label{tab:speed-eval-mrec}
\vspace*{-.5\baselineskip}
\centering
\makebox[\textwidth][c]{
\begin{tabular}{r@{\hskip 0.05in}rr@{\hskip 0.2in}rr@{\hskip 0.05in}r@{\hskip 0.2in}r}
& & \multicolumn{2}{r}{Mathematical (sub)formulae} & \multicolumn{3}{r}{Indexing time (min)} \\
\multicolumn{2}{r}{Documents} & Input & Indexed & \multicolumn{2}{c}{Real (Wall clock)} & CPU \\ \midrule
10,000 & (2.28\,\%) & 3,406,068 & 64,008,762 & 35.75 & (2.05\,\%) & 35.05 \\
100,000 & (22.76\,\%) & 36,328,126 & 670,335,243 & 384.44 & (22.00\,\%) & 366.54 \\
439,423 & (100\,\%) & 158,106,118 & 2,910,314,146 & 1,747.16 & (100\,\%) & 1,623.22 \\
\end{tabular}
}

\medskip

\caption{Quality evaluation results on the NTCIR-11 Math-2 dataset.
The mean average precision (MAP), and precisions at ten (P@10) and five (P@5) are reported for queries formulated using Presentation (PMath), and Content MathML (CMath), a combination of both (PCMath), and \LaTeX. 
Two different relevance judgement levels of $\geq1$ (partially relevant), and $\geq3$ (relevant) were used to compute the measures. Number between slashes (/$\cdot$/) is our rank among all teams of NTCIR-11 Math-2 Task.}
\label{tab:quality-eval-ntcir11}
\centering
\begin{tabular}{llllll}
Measure & Level & PMath & CMath & PCMath & \LaTeX \\ \midrule
  MAP & 3 & 0.3073 & \textbf{0.3630 /1/} & 0.3594 & 0.3357 \\
  P@10 & 3 & 0.3040 & \textbf{0.3520 /1/} & 0.3480 & 0.3380 \\
  P@5 & 3 & 0.5120 & \textbf{0.5680 /1/} & 0.5560 & 0.5400 \\
  P@10 & 1 & 0.5020 & 0.5440 & \textbf{0.5520 /1/} & 0.5400 \\
\end{tabular}
\end{table}

Finally, the users can deploy WebMIaS in a single line of code with the \texttt{dataset} and \texttt{index} directories in a container named \texttt{webmias} running at the TCP port \texttt{8888} on the \texttt{localhost}. 
The WebMIaS system will be running at \url{http://localhost:8888/WebMIaS}.

\section{Evaluation}
\label{sec:3:evaluation}
\vspace*{-.5\baselineskip}

We performed a speed evaluation of MIaS on the MREC dataset~\cite{dml:liska2011short} (see Table~\ref{tab:speed-eval-mrec}), and a quality evaluation on the NTCIR-10~Math~\cite{mir:NTCIR-10-Overview,MIR:MIRMUshort}, NTCIR-11~\mbox{Math-2}~\cite{NTCIR11Math2overviewshort,mir:MIaSNTCIR-11short} (see
Table~\ref{tab:quality-eval-ntcir11}), NTCIR-12~MathIR~\cite{ZanibbiEtAl16NTCIR,RuzickaSojkaLiska16Mathshort}, and ARQMath~2020~\cite{zanibbi2020overview,novotny2020three} datasets. We also measured the time to deploy WebMIaS without Docker (see Figure~\ref{fig:CI}).

The speed evaluation shows that the indexing time of our system is linear in the number of indexed documents and that the average query time is 469\,ms. Additionally, the dockerization of WebMIaS reduces the deployment time from about 10~minutes to a matter of seconds. With respect to quality evaluation, MIaS has notably won the NTCIR-11~Math-2 task.

\section{Conclusion}
\label{sec:4:conclusion}
\vspace*{-.5\baselineskip}

An open-source environment brings reproducibility and the possibility of trying out the projects of one's interest without limitations.
However, the installation instructions are often hard to follow with many prerequisites and possible conflicts with the running operating environment on the go.
Automation tools, continuous integration, and package virtualization ease the development process. 
With this motivation and in the hope of helping the math community, 
we have dockerized our math-aware web search engine WebMIaS.
As a result, anyone can now deploy WebMIaS in a single line of code. 
The software is accessible and at the fingertips of the math community, see \url{https://github.com/MIR-MU/WebMIaS}.

%
%
\bibliographystyle{splncs04}
\enlargethispage*{1\baselineskip}
\vspace*{-.5\baselineskip}
\bibliography{main,sojka}

\begin{thebibliography}{10}
\providecommand{\url}[1]{\texttt{#1}}
\providecommand{\urlprefix}{URL }
\providecommand{\doi}[1]{https://doi.org/#1}

\bibitem{mir:NTCIR-10-Overview}
Aizawa, A., Kohlhase, M., Ounis, I.: {NTCIR-10 Math Pilot Task Overview}. In:
  {Proc. of the 10th NTCIR Conference}. pp. 654--661. NII, Tokyo, Japan (2013)

\bibitem{NTCIR11Math2overviewshort}
Aizawa, A., Kohlhase, M., Ounis, I., Schubotz, M.: {NTCIR-11 Math-2 Task
  Overview}. In: Proc. of the 11th NTCIR Conference. pp. 88--98. NII, Tokyo
  (2014),
  \url{http://research.nii.ac.jp/ntcir/workshop/OnlineProceedings11/pdf/NTCIR/OVERVIEW/01-NTCIR11-OV-MATH-AizawaA.pdf}

\bibitem{boettiger2015introduction}
Boettiger, C.: An introduction to {Docker} for reproducible research. ACM
  SIGOPS Operating Systems Review  \textbf{49}(1),  71--79 (2015)

\bibitem{dml:liska2011short}
L\'{i}\v{s}ka, M., Sojka, P., R\r{u}\v{z}i\v{c}ka, M., Mravec, P.: {Web
  Interface and Collection for Mathematical Retrieval: WebMIaS and MREC}. In:
  Proc. of DML 2011 workshop. pp. 77--84. Masaryk University (2011),
  \url{https://hdl.handle.net/10338.dmlcz/702604}

\bibitem{MIR:MIRMUshort}
Líška, M., Sojka, P., Růžička, M.: {Similarity Search for Mathematics:
  Masaryk University team at the NTCIR-10 Math Task}. In: {Proc.\ of the 10th
  NTCIR Conference}. pp. 686--691. NII, Tokyo, Tokyo (2013),
  \url{https://research.nii.ac.jp/ntcir/workshop/OnlineProceedings10/pdf/NTCIR/MATH/06-NTCIR10-MATH-LiskaM.pdf}

\bibitem{mir:webmias2014short}
Líška, M., Sojka, P., Růžička, M.: Math indexer and searcher web
  interface: Towards fulfillment of mathematicians' information needs. In:
  Proc. of CICM 2014. pp. 444--448. Springer, Zurich (2014),
  \url{https://doi.org/10.1007/978-3-319-08434-3_36}

\bibitem{novotny2020three}
Novotný, V., Sojka, P., Štefánik, M., Lupták, D.: Three is better than one.
  In: CEUR Workshop Proceedings. pp. 1--30. Thessaloniki, Greece (2020),
  \url{http://ceur-ws.org/Vol-2696/paper_235.pdf}

\bibitem{RuzickaSojkaLiska16Mathshort}
R{\r{u}}{\v{z}}i{\v{c}}ka, M., Sojka, P., L{\'i}{\v{s}}ka, M.: {Math Indexer
  and Searcher under the Hood: Fine-tuning Query Expansion and Unification
  Strategies}. In: {Proc. of the 12th NTCIR Conference}. pp. 331--337. NII
  Tokyo (2016),
  \url{https://research.nii.ac.jp/ntcir/workshop/OnlineProceedings12/pdf/ntcir/MathIR/05-NTCIR12-MathIR-RuzickaM.pdf}

\bibitem{mir:MIaSNTCIR-11short}
Růžička, M., Sojka, P., Líška, M.: {Math Indexer and Searcher under the
  Hood: History and Development of a Winning Strategy}. In: Proc. of the 11th
  NTCIR Conference. pp. 127--134 (2014),
  \url{https://is.muni.cz/auth/publication/1201956/en}

\bibitem{ZanibbiEtAl16NTCIR}
Zanibbi, R., Aizawa, A., Kohlhase, M., Ounis, I., Topic, G., Davila, K.:
  {NTCIR}-12 {MathIR} task overview. In: Proc. of the 12th NTCIR. pp. 299--308.
  NII Tokyo (2016)

\bibitem{zanibbi2020overview}
Zanibbi, R., Oard, D.W., Agarwal, A., Mansouri, B.: Overview of {ARQMath} 2020:
  {CLEF} lab on answer retrieval for questions on math. In: Proc. of Int. Conf.
  CLEF 2020. pp. 169--193. Springer (2020)

\end{thebibliography}

\end{document}